\documentclass[onecolumn,secnumarabic,amssymb, nobibnotes, aps, prd]{revtex4-2}

\usepackage{graphicx}
\usepackage{dcolumn}
\usepackage{bm}
\usepackage{amsmath}

\begin{document}

\title{Complete experiment problem for photoproduction of two pseudoscalar mesons on a nucleon in a truncated partial-wave analysis}%

\author{A. Fix}%
\email{fix@tpu.ru}
\affiliation{Tomsk Polytechnic University, Tomsk 634050, Russia}
\author{I. Dementjev}%
\email{iad15@tpu.ru}
\affiliation{Tomsk Polytechnic University, Tomsk 634050, Russia}
\date{}%

\begin{abstract}
It is shown that within a truncated partial wave analysis for photoproduction of two pseudoscalar mesons, it is possible to halve the number of independent partial amplitudes by taking into account parity conservation. In addition, within the framework of the proposed formalism, one can rather easily resolve the double discrete ambiguity arising from the symmetry under complex conjugation of the helicity amplitudes. This leads to essential simplifications of the complete experiment problem, at least as far as mathematical aspects are concerned.\\

\end{abstract}

\maketitle

\section{Introduction}
For single pseudoscalar photoproduction, the necessary and sufficient conditions of a complete experiment for spin amplitudes were
deduced mainly in \cite{BDS} and supplemented in some later works, in
particular, in \cite{Keaton,Chiang}. According to the completeness rules formulated in \cite{BDS}, obtaining the unique solution, including elimination of the discrete ambiguities, requires measurements of not only single, but also four double polarization observables of at least two types. Since it is generally rather hard to accomplish such a task experimentally, another approach has been developed, which may somewhat relax the above requirements.
The case in point is a truncated partial-wave analysis (TPWA) where the partial-wave expansion of the reaction amplitude is truncated at some maximum value of the orbital momentum $L$ (or the total angular momentum $J=L\pm 1/2$).

Significant progress in developing TPWA for single pseudoscalar photoproduction was achieved owing to the results of \cite{Grushin,Omelaenko,Tiator,Wund1,Wund2,Wund3}. In these works not only different complete sets of observables were obtained and systematized (i.e., the mathematical aspect of the problem was elaborated), but also efficient practical methods for extracting multipole amplitudes were devised.

In contrast to single meson photoproduction, analogous methods for two pseudoscalar mesons, like $\gamma N\to \pi\pi N$ or $\gamma N\to \pi\eta N$, are still developed to a much lesser extent. An obvious reason is purely theoretical difficulties arising when one goes to a process with three particles in the final state. In addition to a mere increase of the number of independent kinematic variables, there is a significant growth of the number of amplitudes to be determined.
As a consequence, for example, solution of the complete experiment problem for spin amplitudes requires measuring not only double but also triple polarization observables \cite{Wunder2pi}.

The increasing complexity of the complete experiment for two pseudoscalars is especially crucial for TPWA. As is shown in \cite{FiArTrunc}, the number of independent partial amplitudes with the definite total spin $J$ is equal to $4(2J+1)=8$ for $J=1/2$ and $8(2J+1)$ for higher $J$ values. As one can straightforwardly calculate, the resulting total number of the amplitudes limited by the condition $J\leq J_\mathrm{max}$ is equal to $8(J_\mathrm{max}+1)^2-10$, which already for $J_\mathrm{max}=3/2$ gives 40 amplitudes. Thus, the minimum complete set, providing a unique partial-wave solution up to discrete ambiguities (and up to an overall phase), must contain 79 observables. For comparison, in the single meson case the analogous set for $J_\mathrm{max}=3/2$ requires only 11 observables.

Such a rapid increase of dimension may greatly complicate the complete experiment problem for double meson production. In particular, it may lead to various difficulties of a purely mathematical character in solving a set of nonlinear equations relating observables to bilinear combinations of the amplitudes. These circumstances may substantially weaken the above mentioned advantages of the TPWA method and, thereby, cast doubt on the advisability of its application to the reactions with two mesons.

It is therefore of critical importance that the TPWA problem is formulated in such a way that its dimension is reduced as much as possible.
In the case of a single meson, the parity conservation results in a relationship between different transition amplitudes, what enables one to reduce the number of independent spin amplitudes from 8 to 4. For example, in the helicity basis we have
\begin{equation}\label{aeq1}
A_{-\nu-\lambda-\mu}(\theta)=(-1)^{\nu+\lambda-\mu} A_{\nu\lambda\mu}(\theta)\,,
\end{equation}
where $\lambda$, $\mu$, and $\nu$ are the photon and nucleon helicities in the initial and the final states, $\theta$ is the meson angle in the overall center-of-mass frame. The coordinate system is chosen in such a way that the meson azimuth angle is $\phi=0$. The symmetry (\ref{aeq1}) leads to the corresponding relations for the partial amplitudes $A^J_{\nu\lambda\mu}$ defined by the expansion \cite{Walker}
\begin{equation}\label{aeq2}
A_{\nu\lambda\mu}(\theta)=\sum\limits_J A^J_{\nu\lambda\mu}\
d^J_{\lambda-\mu\nu}(\theta)\,.
\end{equation}
Namely, using the symmetry property of the Wigner rotation matrices $d^J_{-M-M^\prime}(\theta)=(-1)^{M-M^\prime}d^J_{MM^\prime}(\theta)$
one obtains from (\ref{aeq1})
\begin{equation}\label{aeq3}
A^J_{-\nu-\lambda-\mu}=A^J_{\nu\lambda\mu}\,.
\end{equation}

In contrast to single-meson photoproduction, when one goes to a process with two mesons,
the symmetry due to parity conservation reads \cite{Roberts}
\begin{equation}\label{aeq4}
T_{-\nu-\lambda-\mu}(\theta,\phi)=(-1)^{\nu+\lambda-\mu}\,
T_{\nu\lambda\mu}(\theta,2\pi-\phi)\,.
\end{equation}
Here $\phi$ is the angle between the reaction plane and the plane spanned by the momenta of the final nucleon and two mesons. In this case the parity conservation relates the amplitudes $T_{\nu\lambda\mu}$ at different points of the kinematic region and hence does not allow one to reduce their number at one and the same point.

In the present study we show that in contrast to the complete experiment in terms of the spin amplitudes, within TPWA, a rather standard choice of the kinematic variables and the quantization axis makes it possible to obtain the expression like (\ref{aeq3}) relating the corresponding partial wave amplitudes at one and the same point in the reaction phase space. This is achieved by splitting off the $\phi$ dependence of $T_{\nu\lambda\mu}$ in the form of a phase factor. In addition to halving the number of independent amplitudes, the resulting expressions give a method for resolving the double discrete ambiguity which may arise in a partial-wave analysis of the type $S$ observables.

\section{Formalism}

We start from a brief description of the formalism which was outlined in detail in Ref.\,\cite{ArFix}. For definiteness we will use as an example the two pion photoproduction $\gamma N\to\pi\pi N$, although the formulas can be applied to any process in which two pseudoscalar mesons are produced on a nucleon. The reaction formula reads
\begin{equation}\label{aeq5}
\gamma(\omega_\gamma,\vec{k}\,;\lambda)+N(E_i,\vec{p}_i;\mu) \to
\pi(\vec{q}_1,\omega_1)+\pi(\vec{q}_2,\omega_2)+N(E,\vec{p}\,;\nu)\,,
\end{equation}
where in parentheses the four-momenta of the particles and the helicities are given. The transition amplitude in the overall center-of-mass (c.m.) frame is given by a matrix element of the current operator
\begin{equation}\label{aeq6}
T_{\nu\lambda\mu}=-^{(-)}\langle\vec{p},\vec{q}^{\,*};\nu|J_\lambda(\vec{k}\,)|\mu\rangle\,,
\end{equation}
where $\vec{q}^{\,*}$ is the momentum in the $\pi\pi$ c.m.\ frame.
In the nonrelativistic limit it is equal to the relative momentum $(\vec{q}_1-\vec{q}_2)/2$. For the $z$ axis we choose the $\vec{k}$ direction and the $y$ axis is directed along $\vec{p}\times\vec{k}$ (see Fig.1). Now we introduce the multipole expansion of the current
\begin{equation}\label{aeq7}
J_\lambda(\vec{k}\,)=-\sqrt{2\pi}\sum\limits_L i^L\hat{L}\,{\cal O}^{\lambda L}_\lambda(k)\,,
\end{equation}
with ${\cal O}^{\lambda L}_\lambda(k)$ containing the electric and magnetic multipoles
\begin{equation}\label{aeq8}
{\cal O}^{\lambda L}_\lambda(k)=E^L_\lambda+\lambda M^L_\lambda\,,
\end{equation}
and the partial wave expansion of the final state
\begin{eqnarray}\label{aeq9}
&&^{(-)}\langle \vec{p},\vec{q}^{\,*};\nu|=\frac{1}{4\pi}\sum\limits_{l_p j_p m_p}\sum\limits_{l_q m_q}\sum\limits_{JM}\hat{l}_p\hat{l}_q\,
(l_p0\frac12\nu|j_p\nu)
(j_pm_p\, l_q m_q|JM)\,D^{j_p*}_{m_p\nu}(\phi_p,\theta_p,-\phi_p)\\
&&\phantom{xxxxxxx}\times D^{l_q^*}_{m_q 0}(\phi_q,\theta_q,-\phi_q)\,^{(-)}\langle pq^*[(l_p\frac12)j_pl_q]JM|\,,\nonumber
\end{eqnarray}
where the Wigner rotation matrices $D^{j}_{mn}$ are taken in the convention of Rose \cite{Rose}. The notation $\hat{l}$ in Eqs.\,(\ref{aeq7}) and (\ref{aeq9}) stands for $\sqrt{2l+1}$. The angles $(\theta_p,\phi_p=\pi)$ and $(\theta_q,\phi_q)$ are the polar and azimuthal angles of the momenta $\vec{p}$ and $\vec{q}^{\,*}$ in the chosen coordinate frame $OXYZ$.

\begin{figure}[t]
\resizebox{0.48\textwidth}{!}{%
\includegraphics{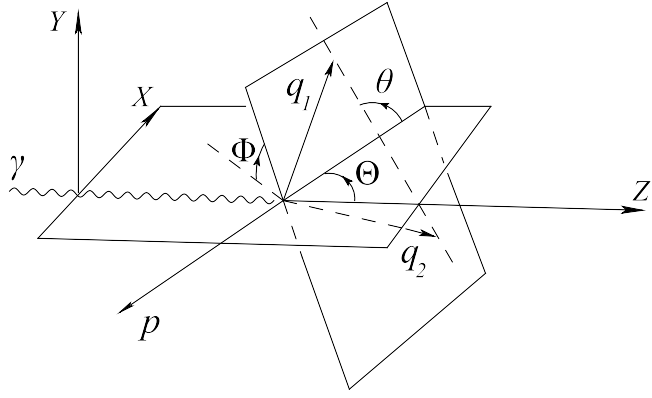}}
\caption{\label{fig1}
Diagram representing the angles $\Theta$, $\theta$ and $\Phi$ used in the present formalism. $\vec{q}_1$, $\vec{q}_2$ and $\vec{p}$ are the three-momenta of the two final mesons and the nucleon, respectively.}
\end{figure}

Now inserting expansions (\ref{aeq7}) and (\ref{aeq9}) into Eq.\,(\ref{aeq6}) and using the Wigner-Eckart theorem one obtains
\begin{widetext}
\begin{eqnarray}\label{aeq10}
&&T_{\nu\lambda\mu}=\frac{1}{2\sqrt{2\pi}}\sum\limits_{l_pj_pm_p}\sum\limits_{l_qm_q}
\sum\limits_{JL}\,i^L(-1)^{l_p+j_p+l_q+J+\nu+\mu}\hat{l}_p\hat{l}_q\hat{j}_p\hat{J}\hat{L}
\left(
\begin{array}{ccc}
  l_p & \frac12 & j_p \\
  0 & \nu & -\nu
\end{array}
\right)
\left(
\begin{array}{ccc}
  j_p & l_q & J \\
  m_p & m_q & \mu-\lambda
\end{array}
\right)
\nonumber\\
&&\phantom{xxxxx}\times
\left(
\begin{array}{ccc}
  J & L & \frac12 \\
  \mu-\lambda & \lambda & -\mu
\end{array}
\right)
\langle pq^*[(l_p\frac12)j_pl_q]J\,||{\cal O}^{\lambda L}||\frac12\,\rangle\,
d^{j_p}_{\nu m_p}(\theta_p)\,D^{l_q*}_{m_q0}(\phi_q,\theta_q,-\phi_q)\,,
\end{eqnarray}
\end{widetext}
where $\langle j'||T^\kappa||j\rangle$ denotes the reduced matrix element. As independent kinematic variables (at fixed photon energy $\omega_\gamma$) we take the polar angle $\Theta$ of the total two-pion momentum $\vec{q}_1+\vec{q}_2$, the $\pi\pi$ invariant c.m.\ energy $\omega_{\pi\pi}$, and the angles $(\theta,\Phi)$. The latter determine orientation of the c.m.\ momentum $\vec{q}^{\,*}$ in the coordinate frame with the $z$-axis along $\vec{q}_1+\vec{q}_2=-\vec{p}$ (see Fig.\,1). The angle $\Phi$ is numerically equal to the azimuth angle $\phi$ in Eq.\,(\ref{aeq4}).

From Fig.\,\ref{fig1} one can see that the set $(\theta_q,\phi_q)$ is related to $(\theta,\Phi)$ via rotation by the angle $\Theta$ around the $OY$ axis. This gives
\begin{equation}\label{aeq11}
D^{l_q*}_{m_q0}(\phi_q,\theta_q,-\phi_q)
=\sum_M
D^{l_q*}_{m_q M}(0,\Theta,0)D^{l_q*}_{M0}(\Phi,\theta,-\Phi)=\sum_M
d^{l_q}_{m_q M}(\Theta)\,e^{iM\Phi}\,d^{l_q}_{M0}(\theta)\,.
\end{equation}
Then taking into account the relation $\theta_p=\pi-\Theta$, leading to
\begin{equation}\label{aeq12}
d^{j_p}_{\nu m_p}(\theta_p)=(-1)^{j_p+\nu}d^{j_p}_{m_p -\nu}(\Theta)\,,
\end{equation}
and using the sum rule for the Wigner matrices
\begin{equation}
\sum\limits_{m_pm_q}
\left(
\begin{array}{ccc}
  j_p & l_q & J \\
  m_p & m_q & \mu-\lambda
\end{array}
\right)
d^{l_q}_{m_q M}(\Theta)\,
d^{j_p}_{m_p -\nu}(\Theta)=(-1)^{M-\nu+\mu-\lambda}\left(
\begin{array}{ccc}
  J & j_p & l_q \\
  \nu-M & -\nu & M
\end{array}
\right)d^J_{\lambda-\mu M-\nu}(\Theta)\,,
\end{equation}
one obtains for the transition amplitude the following expression
\begin{equation}\label{aeq13}
T_{\nu\lambda\mu}(\omega_1,\omega_2,\Theta,\Phi)=\sum\limits_{JM}t^{JM}_{\nu\lambda\mu}(\omega_1,\omega_2)
\,e^{iM\Phi}d^{J}_{\lambda-\mu\, M-\nu}(\Theta)\,.
\end{equation}
Here, instead of $\theta$ and $\omega_{\pi\pi}$, we use as arguments the meson energies $\omega_1$, $\omega_2$. The corresponding relations can readily be derived as Lorentz transformation of these energies
from the $\pi\pi$ to the overall c.m.\ frame:
\begin{equation}\label{aeq14}
\omega_{1/2}=\frac12\Big(W-E\mp\frac{1}{\omega_{\pi\pi}}
\sqrt{E^2-M_N^2}
\sqrt{\omega_{\pi\pi}^2-4m_\pi^2}\,\cos\theta\Big)\,,\quad
E=\frac{1}{2W}\big(W^2-\omega_{\pi\pi}^2+M_N^2\big)\,,
\end{equation}
where $W$ is the total c.m.\ energy.
The partial amplitudes $t^{JM}_{\nu\lambda\mu}$ in Eq.\,(\ref{aeq13}) absorb all quantities which are independent of $\Theta$ and $\Phi$:
\begin{widetext}
\begin{eqnarray}\label{aeq15}
&&t^{JM}_{\nu\lambda\mu}=\frac{1}{2\sqrt{2\pi}}(-1)^{J+M+\nu+\lambda}\sum\limits_{l_pl_qj_pL}
i^L(-1)^{l_p+l_q}\hat{l}_p\hat{l}_q\hat{j}_p\hat{J}\hat{L}
\,d^{l_q}_{M0}(\theta)
\nonumber\\
&&\phantom{xxxxx}\times
\left(
\begin{array}{ccc}
  l_p & \frac12 & j_p \\
  0 & \nu & -\nu
\end{array}
\right)
\left(
\begin{array}{ccc}
  J & L & \frac12 \\
  \mu-\lambda & \lambda & -\mu
\end{array}
\right)
\left(
\begin{array}{ccc}
  J & j_p & l_q \\
  \nu-M & -\nu & M
\end{array}
\right)
\langle pq^*[(l_p\frac12)j_pl_q]J\,||{\cal O}^{\lambda L}||\frac12\,\rangle\,.
\end{eqnarray}
\end{widetext}
It is now an easy matter to establish from (\ref{aeq13}) that
the symmetry analogous to Eq.\,(\ref{aeq4})
\begin{equation}\label{aeq16}
T_{-\nu-\lambda-\mu}(\omega_1,\omega_2,\Theta,\Phi)=(-1)^{\nu+\lambda-\mu}\,T_{\nu\lambda\mu}(\omega_1,\omega_2,\Theta,-\Phi)
\end{equation}
leads to the relation
\begin{equation}\label{aeq17}
t^{J-M}_{-\nu-\lambda-\mu}(\omega_1,\omega_2)=(-1)^M\,t^{JM}_{\nu\lambda\mu}(\omega_1,\omega_2)\,,
\end{equation}
which can also be derived directly from the definition (\ref{aeq15}).

The equations (\ref{aeq13}) and (\ref{aeq17}) are the main results of the present study. As we can see, despite the fact that parity conservation relates the amplitudes $T_{\nu\lambda\mu}$ at different points (Eq.\,(\ref{aeq16})), turning to partial waves allows one to relate the corresponding partial amplitudes $t^{JM}_{\nu\lambda\mu}$ at the same values of $\omega_1,\omega_2$. Specifically, for each $J$ and $M$ we can choose as four independent amplitudes those with $\lambda=1$, i.e.
\begin{equation}\label{aeq18}
t^{JM}_{\frac12 1-\frac12}\,,\quad
t^{JM}_{\frac12 1 \frac12}\,,\quad
t^{JM}_{-\frac12 1-\frac12}\,,\quad
t^{JM}_{-\frac12 1 \frac12}\,.
\end{equation}
It is clear, that the possibility to halve the number of amplitudes in TPWA appears since we are able to isolate the $\Phi$ dependence of $T_{\nu\lambda\mu}$ by the phase factor $e^{iM\Phi}$.

Note that, as one can see from Eq.\,(\ref{aeq15}), the quantum number $M$ has the meaning of $z$-projection of the orbital momentum $l_q$, rather than of the total angular momentum $J$. In the limit $\vec{q}_1\to\vec{q}_2$, only the partial waves with $l_q=0$ and, consequently, $M=0$ contribute. In this situation, the amplitude $T_{\nu\lambda\mu}(\omega_1,\omega_2,\Theta,\Phi)$ does not depend on $\Phi$, so that one has a complete analogy with the single-meson case.
As may be inferred from (\ref{aeq15}), for fixed $J$ and $\nu$, $M$ varies from $\nu-J$ to $\nu+J$. The resulting total number of the amplitudes $t^{JM}_{\nu\lambda\mu}$, taking into account (\ref{aeq17}), is equal to $4(J_\mathrm{max}+1)^2-5$. In particular, for  $J_\mathrm{max}=3/2$, we thus have only 20 amplitudes.

The symmetry (\ref{aeq17}) favorably distinguishes the expansion (\ref{aeq13}) from the partial wave expansion derived in \cite{FiArTrunc}, where the quantization axis was chosen orthogonal to the plane spanned by the momenta of the three final particles. Within this choice, the invariance with respect to parity transformation requires a permutation of the arguments $\omega_1$ and $\omega_2$ (see Eq.\,(8) in Ref.\,\cite{FiArTrunc}) and thus does not allow to get an expression of the form (\ref{aeq17}).

Another advantage of the decomposition (\ref{aeq13}) is that it makes it fairly easy to find discrete ambiguities of TPWA for the group $S$ (single polarization) observables. In the case of photoproduction of two pseudoscalar mesons, in addition to the unpolarized cross section, this group contains 9 observables. Their expressions in terms of helicity amplitudes are listed in Table \ref{tab1}. There we use the shorthand notations
\begin{subequations}\label{eq19}
\begin{gather}
H_1=T_{\frac12 1-\frac12},\quad H_2=T_{\frac12 1 \frac12},\quad
H_3=T_{-\frac12 1-\frac12},\quad H_4=T_{-\frac12 1\frac12},\\
H_5=T_{ \frac12-1-\frac12},\quad H_6=T_{ \frac12-1 \frac12},\quad
H_7=T_{-\frac12-1-\frac12},\quad H_8=T_{-\frac12-1 \frac12}\,.
\end{gather}
\end{subequations}
\begin{table}[ht]
\renewcommand{\arraystretch}{1.7}
\centering \caption{Single polarization observables for photoproduction of two pseudoscalar mesons in terms of helicity amplitudes. The notations in the first column are taken from Ref.\,\cite{ArFix}. The observable $I$ is determined by the unpolarized differential cross section as $d\sigma/d\Omega={\cal K}I$ with ${\cal K}$ being the reaction kinematic factor. The symbolic differential $d\Omega$ is determined as $d\Omega\equiv d\omega_1\,d\omega_2\,d\cos\Theta\,d\Phi$. The notation $\hat{\cal O}$ means the polarized cross section ${\cal O}I$. In the third column, we also indicate the notations introduced by Roberts and Oed in Ref.\,\cite{Roberts}, which are commonly used in the literature.
The last column contains the corresponding observables for single pseudoscalar photoproduction.
}
\begin{tabular}{@{\hspace{0.3cm}}l@{\hspace{0.5cm}}c@{\hspace{0.9cm}}l@{\hspace{0.7cm}}c@{\hspace{0.5cm}}}
\hline \hline Observable & Helicity representation & \cite{Roberts} & $\gamma N\to\pi N$ \\
\hline
$I$ & $ \frac14\,\sum\limits_{i=1}^8|H_i|^2 $ & $I_0$ & $I$ \\
$ \hat{T}^l_{00} $ & $ -\frac12\,\mathrm{Re}\big[
H^*_1H_5+H^*_2H_6+H^*_3H_7+H^*_4H_8\big] $ & $I_0I^c$ & $\hat{\Sigma}$ \\
$ \hat{S}^0_{11} $ & $ -\frac{1}{\sqrt{2}}\,\mathrm{Im}\big[
H^*_2H_1+H^*_4H_3+H^*_6H_5+H^*_8H_7\big] $ & $I_0P_y$ & $\hat{T}$ \\
$ \hat{P}^0_y $ & $ \frac12\,\mathrm{Im}\big[
H^*_1H_3+H^*_2H_4+H^*_5H_7+H^*_6H_8\big] $ & $I_0P_{y^\prime}$ & $\hat{P}$
\\
\hline
$ \hat{T}^c_{00} $ & $ \frac14\,\big[|H_1|^2+|H_2|^2+|H_3|^2+|H_4|^2
-|H_5|^2-|H_6|^2-|H_7|^2-|H_8|^2\big] $ & $I_0I^\odot$ &  \\
$ \hat{S}^l_{00} $ & $ -\frac12\,\mathrm{Im}\big[
H^*_1H_5+H^*_2H_6+H^*_3H_7+H^*_4H_8\big] $ & $I_0I^s$ &  \\
$ \hat{T}^0_{11} $ & $ -\frac{1}{\sqrt{2}}\,\mathrm{Re}\big[
H^*_2H_1+H^*_4H_3+H^*_6H_5+H^*_8H_7\big] $ & $I_0P_x$ &  \\
$ \hat{T}^0_{10} $ & $ \frac14\,\big[|H_1|^2-|H_2|^2+|H_3|^2-|H_4|^2
+|H_5|^2-|H_6|^2+|H_7|^2-|H_8|^2\big] $ & $I_0P_z$ &  \\
$ \hat{P}^0_x $ & $ \frac12\,\mathrm{Re}\big[
H^*_1H_3+H^*_2H_4+H^*_5H_7+H^*_6H_8\big] $ & $I_0P_{x^\prime}$ &  \\
$ \hat{P}^0_z $ & $ \frac14\,\big[|H_1|^2+|H_2|^2-|H_3|^2-|H_4|^2
+|H_5|^2+|H_6|^2-|H_7|^2-|H_8|^2\big] $ & $I_0P_{z^\prime}$ &  \\
\hline
\hline
\end{tabular}
\label{tab1}
\end{table}
The first four observables
\begin{equation}\label{aeq20}
I\,,\quad \hat{T}^l_{00}\,,\quad
\hat{S}^0_{11}\,,\quad \hat{P}^0_y
\end{equation}
are equivalent, respectively, to the differential cross section, beam, target, and recoil polarization for single pion photoproduction (last column in Table\,\ref{tab1}). In particular, in the coplanar kinematics ($\Phi=0$) we have from (\ref{aeq16})
\begin{equation}\label{aeq20a}
H_5=H_4,\quad H_6=-H_3,\quad H_7=-H_2,\quad H_8=H_1,
\end{equation}
and the expressions for the observables (\ref{aeq20}) formally coincide with those in the single meson case (for $\Sigma$ and $T$ up to numerical factors $-1$ and $-\sqrt{2}$, due to differences in the definitions). In the same limit the remaining six observables in Table\,\ref{tab1} vanish exactly.

Using (\ref{aeq13}), one can obtain the corresponding expansions of the observables in a general form
\begin{equation}\label{aeq21}
{\cal O}^\alpha(\Theta,\,\Phi,\,\omega_1,\,\omega_2)=\mathrm{Re}/\mathrm{Im}\Big[\sum_{jm^\prime m}e^{-im\Phi}d^j_{m^\prime m-M_\alpha}(\Theta)\,u^\alpha_{jm^\prime m}(\omega_1,\,\omega_2)\Big]\,.
\end{equation}
Here $M_\alpha$ depends on the type of the observable ${\cal O}^\alpha$ (it is equal to $+1$ or $-1$ if the final nucleon is polarized and is $0$ otherwise). The properties of the expansion coefficients $u^\alpha_{jm^\prime m}$ will be considered in more detail elsewhere. Here we only note that for the set (\ref{aeq20}) the truncation in (\ref{aeq13}) at $J_\mathrm{max}=3/2$ gives 53 independent coefficients $u^\alpha_{jm^\prime m}$ at each point $(\omega_1,\omega_2)$ of the Dalitz plot. Of these coefficients, one can always choose 39 independent ones, which allow fixing all 20 complex amplitudes $t^{JM}_{\nu\lambda\mu}$
up to a common phase (and up to possible discrete ambiguities).

Using the expansion (\ref{aeq13}) it is easy to verify that the transformation
\begin{equation}\label{aeq22}
t^{JM}_{\frac12 1-\frac12}\to t^{JM*}_{\frac12 1-\frac12}\,,\quad
t^{JM}_{\frac12 1 \frac12}\to -t^{JM*}_{\frac12 1\frac12}\,,\quad
t^{JM}_{-\frac12 1-\frac12}\to -t^{JM*}_{-\frac12 1-\frac12}\,,\quad
t^{JM}_{-\frac12 1 \frac12}\to t^{JM*}_{-\frac12 1 \frac12}
\end{equation}
results in the following transformation of the helicity amplitudes
\begin{equation}\label{aeq23}
H_1\leftrightarrow H_8^*\,,\quad
H_2\leftrightarrow H_7^*\,,\quad
H_3\leftrightarrow H_6^*\,,\quad
H_4\leftrightarrow H_5^*\,,
\end{equation}
which, as one can see directly from Table \ref{tab1}, leaves the observables (\ref{aeq20}) unchanged. In this case $\hat{S}^l_{00}$, $\hat{T}^0_{11}$, and $\hat{P}^0_x$ also remain invariant, whereas
$\hat{T}^c_{00}$, $\hat{T}^0_{10}$, and $\hat{P}^0_z$ change sign.

The ambiguity (\ref{aeq22}) is analogous to the double discrete ambiguity in single pseudoscalar meson photoproduction considered in Ref.\,\cite{FixDem}.   In particular, in the limit $\Phi=0$, when the relations (\ref{aeq20a}) hold, replacement (\ref{aeq22}) results in the transformation
\begin{equation}\label{aeq24}
H_1\to  H_1^*\,,\quad
H_2\to -H_2^*\,,\quad
H_3\to -H_3^*\,,\quad
H_4\to  H_4^*\,,
\end{equation}
closely resembling the well-known discrete ambiguity existing in the single meson case \cite{Keaton,Chiang}.

\section{Conclusion}
We developed the formalism for photoproduction of two pseudoscalar mesons, which enables one to halve the number of independent partial amplitudes by taking into account parity conservation. In addition, this formalism
facilitates a study of the double discrete ambiguity arising from the symmetry under complex conjugation of the helicity amplitudes. These simplifications make the corresponding complete experiment problem easier, at least as far as its mathematical aspects are concerned.

Since the number of independent amplitudes grows rapidly with increasing $J_\mathrm{max}$, studying the cases $J_\mathrm{max}>3/2$ can hardly be of  any practical significance. At the same time, $J_\mathrm{max}=3/2$ discussed here is, apparently, the minimal nontrivial case for photoproduction of two pseudoscalars which may be related to the real situation
(the simplest case $J_\mathrm{max}=1 /2$ which is more of a pedagogical use
is considered in detail in \cite{FiArTrunc}). In particular, according to most if not all of the $\gamma N\to\pi^0\pi^0 N$ analyzes \cite{Laget,Oset,Ochi,Prakh2pi,Sarantsev},
at the energies below the $N(1520)3/2^-$ resonance, the partial waves with $J\leq 3/2$ dominate the reaction amplitude. Here a truncated PWA would be especially useful for revealing the mechanisms which are responsible for the linear increase of the $\gamma N\to\pi^0\pi^0 N$ total cross section.

Another question which may arise in connection with our results is whether (\ref{aeq22}) is the only discrete transformation of the amplitudes $t^{JM}_{\nu\lambda\mu}$ which leaves the group (\ref{aeq20}) unchanged? Isn't there a wider symmetry which applies to other type $S$ observables in addition to (\ref{aeq20}), or even to the entire set listed in Table\,\ref{tab1}? Based on the reasoning that any such transformation generates a corresponding transformation of the helicity amplitudes $H_i$, $i=1,\,\ldots,\,8,$ which at $\Phi=0$ should reduce to (\ref{aeq24}), the symmetry (\ref{aeq22}) is indeed unique. This issue, however, requires a more detailed study.

\begin{acknowledgements}
Financial support for this work was provided by the Russian Science Foundation, grant No. 22-42-04401.
\end{acknowledgements}

\end{document}